\documentclass[%
 reprint,
 amsmath,amssymb,
 aps,
]{revtex4-1}

\usepackage{graphicx}
\usepackage{dcolumn}
\usepackage{bm}

\begin{document}

\preprint{APS/123-QED}

\title{Experimental generation of multiple quantum correlated beams from hot rubidium vapor}

\author{Zhongzhong Qin,$^{1}$ Leiming Cao,$^{1}$ Hailong Wang,$^{1}$ A. M. Marino,$^{2}$ Weiping Zhang,$^{1}$ and Jietai Jing$^{1,*}$} 
\affiliation{%
$^1$State Key Laboratory of Precision Spectroscopy, Quantum Institute for Light and Atoms, Department of Physics, East China Normal University, Shanghai 200062, China\\
$^2$Homer L. Dodge Department of Physics and Astronomy, The University of Oklahoma, 440 W. Brooks St., Norman, Oklahoma 73019, USA.\\
}%

\begin{abstract}
Quantum correlations and entanglement shared among multiple quantum modes are important for both fundamental science and the future development of quantum technologies. This development will also require an efficient quantum interface between multimode quantum light sources and atomic ensembles, which makes it necessary to implement multimode quantum light sources that match the atomic transitions. Here we report on such a source that provides a method for generating quantum correlated beams that can be extended to a large number of modes by using multiple four-wave mixing (FWM) processes in hot rubidium vapor. Experimentally we show that two cascaded FWM processes produce strong quantum correlations between three bright beams but not between any two of them. In addition, the intensity-difference squeezing is enhanced with the cascaded system to -7.0 $\pm$ 0.1 dB from the -5.5 $\pm$ 0.1/-4.5 $\pm$ 0.1 dB squeezing obtained with only one FWM process. One of the main advantages of our system is that as the number of quantum modes increases, so does the total degree of quantum correlations. The proposed method is also immune to phase instabilities due to its phase insensitive nature, can easily be extended to multiple modes, and has potential applications in the production of multiple quantum correlated images.
\end{abstract}

\pacs{Valid PACS appear here}
\maketitle


Multipartite entanglement and correlations have attracted considerable attention because of their fundamental scientific significance \cite{EPR,GHZ} and potential applications in future quantum technologies \cite{LoockRMP,Kimble}. This is particularly true in optics since light is an ideal candidate as a carrier of information \cite{LukinRMP}. Significant progress has been made in this field with the experimental demonstration of topological error correction with an eight-photon cluster state in the discrete variable regime \cite{Pan2012} and the generation of eight entangled modes in the continuous-variable (CV) regime using a programmable virtual network \cite{PKLam2012}.

A number of different techniques for the generation of entanglement between multiple beams of light have been proposed and experimentally implemented. For example, in the CV regime, three-beam entanglement has been generated between the signal, idler, and pump beams of an optical parametric oscillator (OPO) \cite{Villar2009} and by using two cascaded OPOs \cite{JiaPRL}. In the discrete variable regime, cascaded spontaneous parametric down-conversion has been used to generate quantum correlated photon triplets \cite{tripletNature} and three-photon energy-time entanglement \cite{tripletNP}.

Here we focus on the CV regime, which offers the advantages of unconditional quantum state generation and high efficiency quantum detection. The standard technique for generating CV entanglement between multiple beams is based on mixing squeezed states on linear beamsplitters \cite{Loock2000,JingPRL,Loock2003,Furusawa}. As a result of its application for one-way quantum computing, another type of CV multipartite entangled state, the cluster state, has been theoretically proposed \cite{Zhang2006,Nielsen} and experimentally demonstrated \cite{Su2007}. The main limitation with these techniques is that as the number of modes increases, optical losses, mode mismatch, and the required phase stability between the different modes degrade the quantum correlations and limit the maximum number of entangled beams.

A promising alternative to these schemes is to produce one or two quantum states of light composed of multiple modes.  This has been experimentally achieved with combinations of different spatial regions of one beam \cite{PKLam2012}, multiple longitudinal modes \cite{Olivier2011}, or temporal modes from OPO \cite{Treps}. Recently, it has also been experimentally demonstrated that a pair of multimode intensity-correlated beams \cite{PaulOL,LiuOL,QinOL} and quantum entangled images \cite{PaulScience} can be successfully produced with a four-wave mixing (FWM) process in hot rubidium vapor. This system has proven to be very successful for a number of applications, such as the tunable delay of EPR entangled states \cite{PaulNature}, the realization of a SU(1,1) nonlinear interferometer \cite{JingAPL,KongAPL}, and the generation of high purity narrowband single photons \cite{Lvovsky}.  Although these systems have the advantage of being able to produce quantum states of lights with many modes, all the modes are in a single beam, making it hard to address them independently.

The main experimental limitation in previous schemes is generating multimode quantum states with a high degree of quantum correlations in a way that can be extended to a large number of beams. In this paper, we theoretically propose and experimentally demonstrate a method that overcomes this limitation.  The method is based on cascaded FWM processes in an atomic system.  The use of atomic systems offers the additional advantage of producing spatially multimode quantum states of light that are naturally matched to an atomic transition.  This makes them ideal for use in quantum manipulation that requires high efficiency mapping, storage and retrieval of quantum states of light in and out of an atomic medium \cite{PanPNAS,Rempe,PaulNature,Lvovskymemory}.

Our scheme uses one of the most popular candidates for the generation of quantum correlated twin beams, a parametric amplifier (PA) as shown in Fig. 1(a). A coherent probe beam (Pr$_{0}$) with an intensity $I_0$ is seeded into a PA (PA$_{1}$), with a gain of $G$, where it crosses with a pump beam (P$_{1}$). The output probe beam is amplified (Pr$_{1}$) and a conjugate beam (C$_{1}$) is simultaneously generated. The intensities of these twin beams (Pr$_{1}$ and C$_{1}$) are $I'_0=GI_0$ and $I_1=(G-1)I_0$ respectively. Although the total power of the twin beams is significantly amplified, the variance of the relative intensity difference between them remains unchanged after the amplification. As a result, the relative intensity difference of beams Pr$_{1}$ and C$_{1}$ is squeezed compared with the corresponding shot noise limit (SNL) by an amount of $1/(2G-1)$. We then pick out one of the twin beams (say Pr$_{1}$ as shown in Fig. 1(b)) and use it to seed a second identical PA (PA$_{2}$). This beam is amplified (Pr$_{2}$) with a gain of $G$, and at the same time, a new conjugate beam (C$_{2}$) is generated. The intensities of these two newly generated twin beams (Pr$_{2}$ and C$_{2}$) are $I_3=G^{2}I_0$, and $I_2=G(G-1)I_0$ respectively. If one calculates the intensity-difference noise of the three generated beams (C$_{1}$, C$_{2}$ and Pr$_{2}$) given by $I_3-I_2-I_1$ and compares it with the corresponding SNL, one will find that the degree of intensity-difference squeezing of the triple beams is given by $1/(2G^{2}-1)$. If we extend this system to a series of PAs, as shown in Fig. 1(c), then we would get one amplified probe beam (Pr$_{n}$) and $n$ newly generated conjugate beams (C$_{1}$, C$_{2}$, \dots, C$_{n}$), where $n$ is the number of the PAs. The amount of intensity-difference squeezing of the $n+1$ quantum correlated beams is given by $1/(2G^{n}-1)$. One can see that the amount of squeezing present in this system increases as the number of quantum modes increases. In other words, we can enhance the quantum correlations in our system by increasing the number of quantum modes. Another advantage of our system is the phase-insensitivity that makes it possible to easily extend our system to a large number of modes, as it does not require relative phase stability between all the parametric amplification processes.
\begin{figure}
\includegraphics[width=7cm]{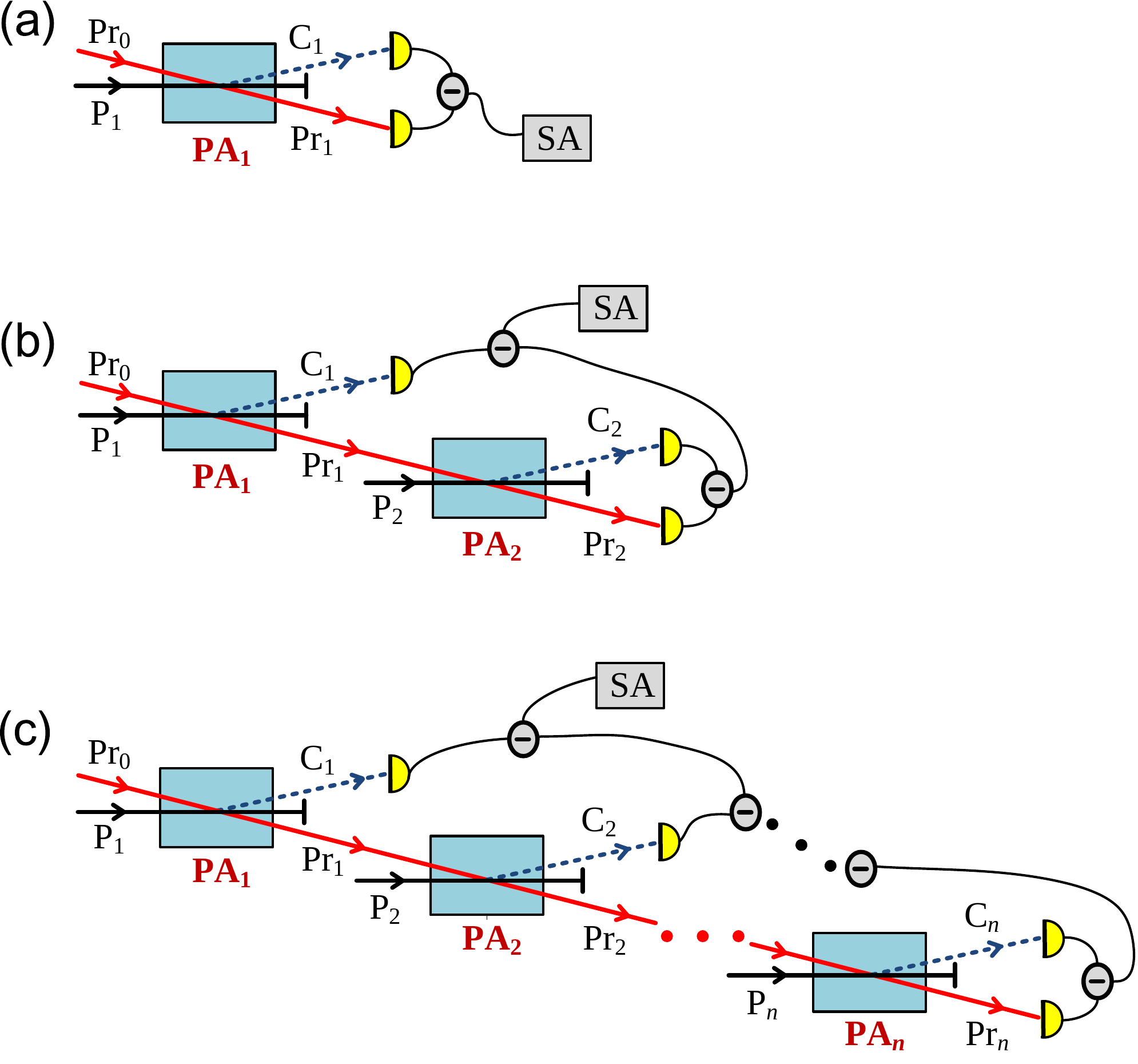}
\caption{\label{fig:epsart} (color online). Proposed system for the generation of multiple quantum correlated beams. (a) Single PA configuration for generating quantum correlated twin beams; (b) Cascaded two-PA configuration for generating quantum correlated triple beams; (c) Cascaded $n$-PA configuration for generating $n+1$ quantum correlated beams. PA$_{i}$, the ith parametric amplifier; P$_{i}$, the ith pump beam; Pr$_{i}$, the ith probe beam, with Pr$_{0}$ as the initial probe beam; C$_{i}$, the ith conjugate beam; SA, spectrum analyzer.}
\end{figure}

\begin{figure*}
\includegraphics[width=12cm]{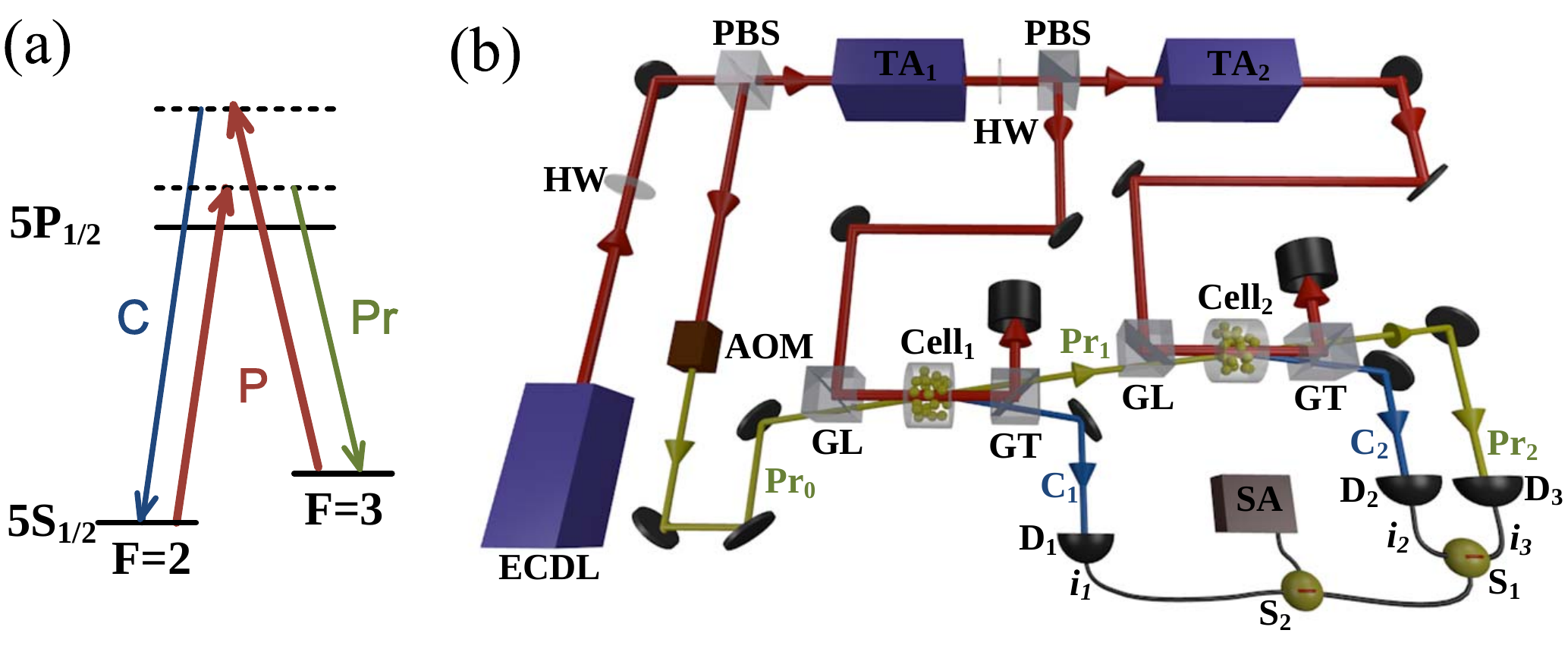}
\caption{\label{fig:wide} (color online). Experimental layout for generating and detecting quantum correlated triple beams. (a) Double-$\Lambda$ scheme in the D1 line of $^{85}$Rb. P: pump; Pr: probe, C: conjugate. (b) Experimental setup. ECDL, external cavity diode laser; HW, half waveplate; PBS, polarization beam splitter; TA$_{1}$, TA$_{2}$, tapered amplifiers; AOM, acousto-optic modulator; GL, Glan-Laser polarizer; GT, Glan-Thompson polarizer; Cell$_{1}$, Cell$_{2}$, first and second rubidium vapor cell; D$_{1}$, D$_{2}$, D$_{3}$, photodetectors; S$_{1}$, S$_{2}$, subtractors; SA, spectrum analyzer; Pr$_{0}$, initial probe beam; Pr$_{1}$, Pr$_{2}$, first and second probe beam; C$_{1}$ and C$_{2}$, first and second conjugate beam.}
\end{figure*}

Our experimental layout is shown in Fig. 2. The two PAs are based on a FWM process in a double-$\Lambda$ configuration in a $^{85}$Rb vapor cell (Fig. 2a). We use an external cavity diode laser (ECDL) and two diode laser tapered amplifiers (TAs) as our laser system. The ECDL has a linewidth of 100 kHz tuned about 0.8 GHz to the blue of the $^{85}$Rb $5S_{1/2}, F=2\rightarrow5P_{1/2}$ transition with a total power of around 90 mW. A polarizing beam splitter (PBS) is used to split the beam into two. One of the beams goes through the two TAs (TA$_{1}$, TA$_{2}$) in series to generate the two pump beams needed for the experiment. The other beam is double-passed through an acousto-optic modulator (AOM). In this way, a much weaker probe beam tuned 3.04 GHz to the red of the pump is derived, which results in very good relative frequency stability of the initial probe beam with respect to the pump beams.

The two Rb vapor cells are 12.5 mm long and temperature-stabilized at around 110 $^{\circ}$C and 108 $^{\circ}$C respectively. They are illuminated by intense vertically polarized pump beams (both Pump$_{1}$ and Pump$_{2}$ are about 210 mW) with beam waists of about 550 $\mu$m. A horizontally polarized probe beam (Pr$_{0}$) with a power of about 20 $\mu$W and beam waist of 280 $\mu$m is combined with beam Pump$_{1}$ at an angle of 6 mrad at the center of Cell$_{1}$ by a Glan-Laser polarizer. A Glan-Thompson polarizer with an extinction ratio of 10$^{5}$:1 at the output port of the vapor cell is used to filter out the pump beam. Based on these settings, the initial probe beam (Pr$_{0}$) is amplified by a gain of $G_1\approx2.9$, becoming Pr$_{1}$. At the same time, a conjugate beam (C$_{1}$) with a frequency of 3.04 GHz blueshifted from the pump is produced by the FWM process. As a result, the beams Pr$_{1}$ and C$_{1}$ have powers of about 58 $\mu$W and 41 $\mu$W respectively. The beam C$_{1}$ is then picked out for direct detection with a photodetector (D$_{1}$). By using a 4f imaging system, we map the beam Pr$_{1}$ from the center of Cell$_{1}$ to the center of Cell$_{2}$. It is then combined with beam Pump$_{2}$ at the exact same angle (6 mrad) at the center of Cell$_{2}$ with another Glan-Laser polarizer. Beam Pr$_{1}$ is amplified by a gain of $G_2\approx2.1$ in the second FWM process, becoming Pr$_{2}$, and a new conjugate beam (C$_{2}$) is generated at the same time. After considering the imperfect optics used between the two cells, which results in 7\% loss for Pr$_{1}$, the powers of Pr$_{2}$ and C$_{2}$ are about 115 $\mu$W and 71 $\mu$W respectively. After Cell$_{2}$, the two newly generated beams (Pr$_{2}$ and C$_{2}$) are sent to two photodiodes (D$_{3}$ and D$_{2}$) respectively. The detectors's transimpedance gain is 10$^{4}$ V/A  and quantum efficiency is 96\%. The obtained photocurrents \emph{i$_{1}$}, \emph{i$_{2}$}, \emph{i$_{3}$} are analyzed by two methods. On one hand, they are directly sent to a digital oscilloscope (not shown in Fig. 2) to investigate their temporal waveform correlations (see Supplemental Material). On the other hand, they are subtracted from each other in the form of $i_3-i_2-i_1$ by using two radiofrequency subtractors (S$_{1}$, S$_{2}$) and then analyzed with a spectrum analyzer (SA).

To verify the predicted strong quantum correlations in our system, we measure the noise power spectra of the three photocurrents \emph{i$_{1}$}, \emph{i$_{2}$}, \emph{i$_{3}$} (indicated as Trace \emph{A}, \emph{B}, \emph{C} respectively) and their subtractions $i_3-i_2$, $i_3-i_1$, $i_2-i_1$, $i_3-i_2-i_1$ (indicated as Trace \emph{D}, \emph{E}, \emph{F} and \emph{G} respectively) with a SA set to a 30 kHz resolution bandwidth (RBW) and a 300 Hz video bandwidth (VBW). This gives the variances of these photocurrents. The results are shown in Fig. 3(a). All of these seven traces are normalized to the corresponding SNLs (Trace \emph{H}). The red straight line at 0 dB is taken as a reference, which corresponds to the average value of data points on Trace \emph{H}. We calibrate the SNL of the triple beams by using a beam in a coherent state with a power equal to the total power of the triple beams impinging on the photodetectors. We then split it with a 50/50 beam splitter, direct the obtained beams into two of the photodiodes, D$_{2}$ and D$_{3}$, and record the noise power of the difference of the photocurrents. This balanced detection system makes it possible to cancel all the sources of classical noise and obtain a measure of the SNL and is equivalent to performing a balanced homodyne detection of the vacuum.  It thus provides an accurate measure of the SNL.

\begin{figure}[b]
\includegraphics[width=8.5cm]{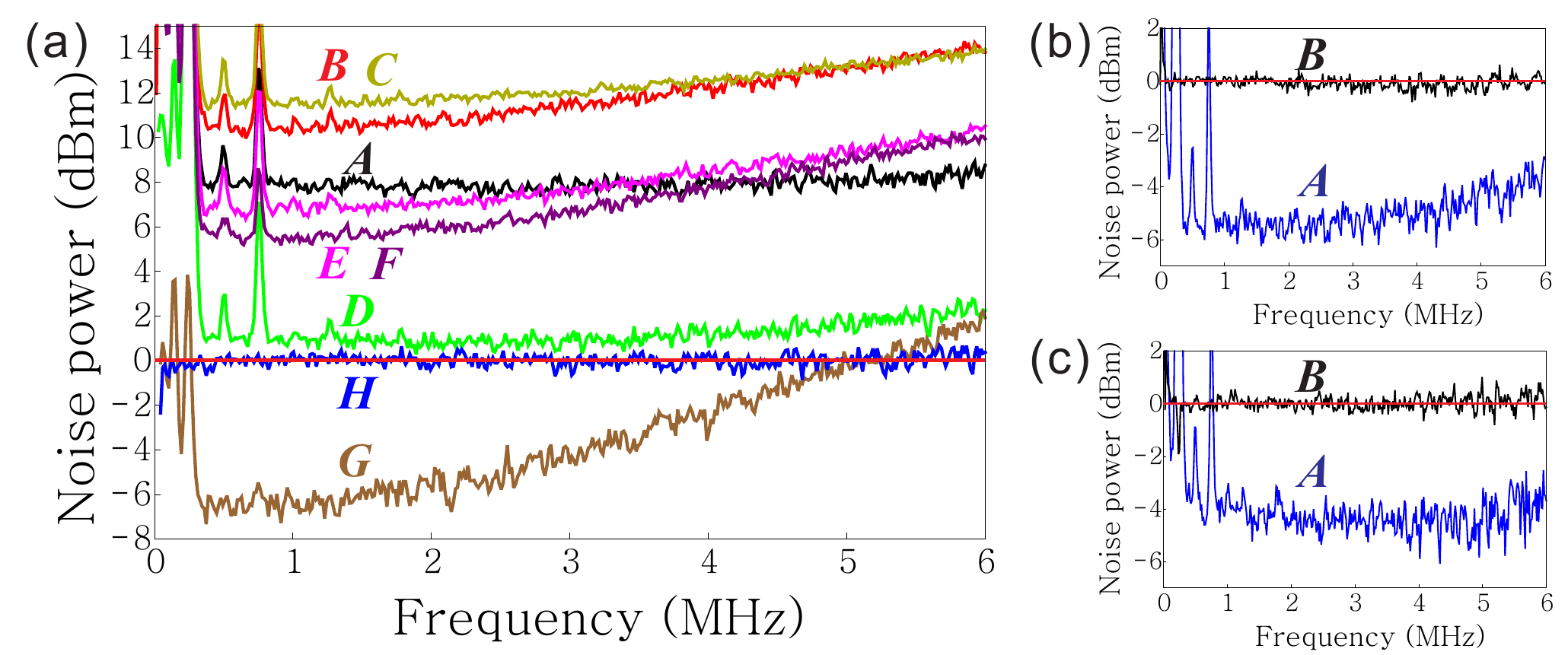}
\caption{\label{fig:epsart} (color online). Quantum correlations of the triple beams. (a) Normalized noise power of (\emph{A}) \emph{i$_{1}$}; (\emph{B}) \emph{i$_{2}$}; (\emph{C}) \emph{i$_{3}$}; (\emph{D}) $i_3-i_2$; (\emph{E}) $i_3-i_1$; (\emph{F}) $i_2-i_1$; (\emph{G}) $i_3-i_2-i_1$; (\emph{H}) the corresponding SNLs of Trace \emph{A} $\sim$ \emph{G}; (b) Normalized intensity-difference noise power of the twin beams generated from Cell$_{1}$ (\emph{A}) and the corresponding SNL (\emph{B}); (c) Normalized intensity-difference noise power of the twin beams generated from Cell$_{2}$ (\emph{A}) and the corresponding SNL (\emph{B}). The electronic noise floor and background noise are all about 6 dB below the corresponding SNLs at 1 MHz and have been subtracted from all of the traces.}
\end{figure}

We first record the photocurrent noise power of \emph{i$_{1}$}, \emph{i$_{2}$} and \emph{i$_{3}$} and we find that they are all above their corresponding SNLs. Trace \emph{A} is around 8 dB above the corresponding SNL, because the noise of beam C$_{1}$ is amplified in the first FWM process. Trace \emph{B} and Trace \emph{C} are around 11 dB above the corresponding SNLs, because the noise of beams C$_{2}$ and Pr$_{2}$ is amplified twice in the two FWM processes.

We then investigate the pairwise intensity correlations for any pair of the three photocurrents. We subtract \emph{i$_{2}$} from \emph{i$_{3}$} and record the noise power, which gives the intensity-difference noise of beams Pr$_{2}$ and C$_{2}$ ($i_3-i_2$). Trace \emph{D} is above its corresponding SNL and clearly shows no intensity-difference squeezing. It is not difficult to prove that the normalized intensity-difference noise level of these two beams is given by $(2G_{1}-1)/(2G_{2}-1)$. So there is no intensity-difference squeezing between them when $G_1>G_2$ (in our case, $G_1\approx2.9$, $G_2\approx2.1$), which agrees with Trace \emph{D}. The other two pairwise intensity-difference noise spectra for $i_3-i_1$ and $i_2-i_1$ are given by Trace \emph{E} and \emph{F} respectively. They are also well above their corresponding SNLs and clearly show no squeezing.

\begin{figure}[b]
\includegraphics[width=8.5cm]{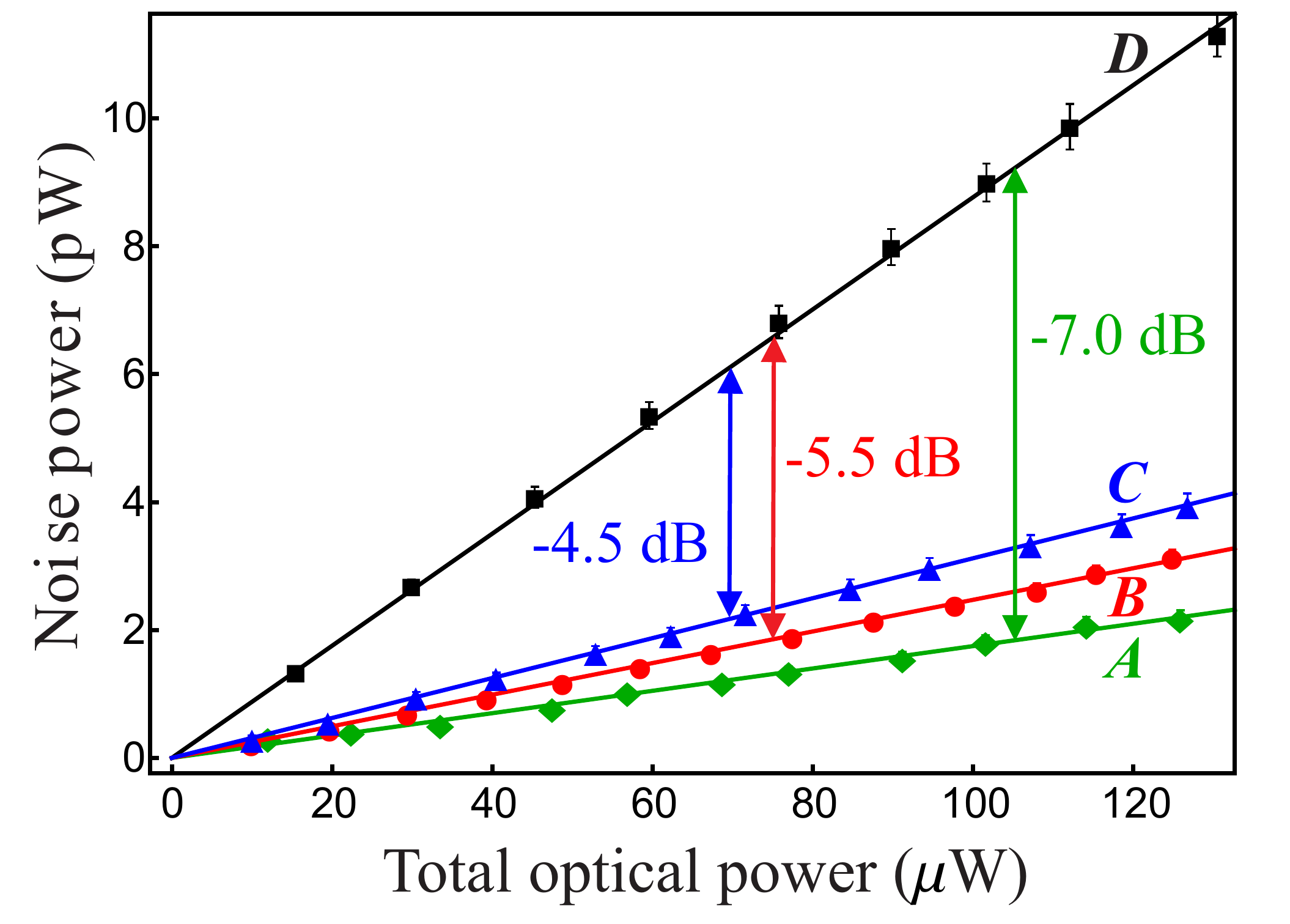}
\caption{\label{fig:epsart} (color online). Enhancement of quantum correlation. Relative intensity noise power at different total optical power for (\emph{A}) triple beams from the cascaded configuration (green diamonds); (\emph{B}) twin beams from Cell$_{1}$ (red circles); (\emph{C}) twin beams from Cell$_{2}$ (blue triangles) and (\emph{D}) SNL (black squares). All these four noise power curves fit to straight lines. The electronic noise floor and background noise are subtracted from all of the data points.}
\end{figure}

Most interesting is Trace \emph{G}, which gives the normalized intensity-difference noise of the triple beams. To get this trace we subtract \emph{i$_{1}$} from $i_3-i_2$ and record the noise power. In principle, the noise power can get back to the noise level of the initial probe beam (Pr$_{0}$). But due to optical losses, absorption from the rubidium vapor and imperfect quantum efficiency (96\%) of the photodetectors, the photocurrent variance of $i_3-i_2-i_1$ is slightly higher than the one of the input probe beam (Pr$_{0}$) (not shown in this figure). As we can see, the intensity-difference noise power of the triple beams has a minimum of 6.5 $\pm$ 0.4 dB below the SNL under these experimental conditions. The large peaks shown below 1 MHz are classical noise from our lasers. These peaks are eliminated almost perfectly on Trace \emph{G}, which shows good noise cancellation in our balanced detection system. The presence of intensity-difference squeezing only among the three beams but not between any two of them shows the tripartite nature of the quantum correlations produced by the cascaded FWM processes.

As shown in Fig. 3(b) and (c), the maximal degrees of intensity-difference squeezing of twin beams from Cell$_{1}$ and Cell$_{2}$ are around -5.4 $\pm$ 0.4 dB and -4.4 $\pm$ 0.4 dB respectively. Trace \emph{A} and \emph{B} are the intensity-difference noise of twin beams from Cell$_{1}$ (Fig. 3(b)) and Cell$_{2}$ (Fig. 3(c)) and the corresponding SNL.

To better show the squeezing enhancement as predicted by the theory, we measure the relative intensity noise power for the triple beams from the cascaded configuration (Curve \emph{A} in Fig. 4) and twin beams from a single cell (Curve \emph{B} and \emph{C} for Cell$_{1}$ and Cell$_{2}$ respectively) at 1 MHz as a function of the total optical power impinging on the photodetectors. Similarly, we also record the noise power of a coherent beam at different optical power using the SNL measurement method described above (Curve \emph{D}). After fitting all these four noise power curves to straight lines, we find that the ratios of slopes between Curve \emph{B}/\emph{C} and Curve \emph{D} are equal to 0.282 $\pm$ 0.003 and 0.356 $\pm$ 0.003 respectively, which shows that the degrees of intensity-difference squeezing of the twin beams from Cell$_{1}$ and Cell$_{2}$ are about -5.5 $\pm$ 0.1 dB and -4.5 $\pm$ 0.1 dB respectively. The ratio of slopes between Curve \emph{A} and Curve \emph{D} is equal to 0.199 $\pm$ 0.003, which shows that the degree of intensity-difference squeezing of the triple beams is enhanced to about -7.0 $\pm$ 0.1 dB (See Fig. 4). The FWM on which our method is based has been shown to operate very close to the quantum limit \cite{Pooser,MarinoPRA}; as such, it can be made to operate without a significant amount of excess noise. Thus, it is possible to obtain large amounts of squeezing with this process and observe an increase in the level of squeezing after the second process in our system.

In conclusion, we have demonstrated a cascadable technique to create and measure quantum correlations among multiple beams produced by multiple FWM processes in hot rubidium vapor. We have shown that quantum squeezing exists between the three beams but not between any two of them when two cascaded FWM processes are used. Compared with the degree of intensity-difference squeezing of the twin beams obtained with a single cell, the degree of intensity-difference squeezing of the triple beams has been enhanced from -5.5 $\pm$ 0.1/-4.5 $\pm$ 0.1 dB to -7.0 $\pm$ 0.1 dB.  In this sense, our method for generating multimode quantum states offers significant advantages over other methods since the quantum correlations increase as the number of quantum modes increases. Compared to the linear beam-splitting method \cite{Loock2000,JingPRL,Loock2003,Furusawa}, our method can compensate or even enhance the quantum correlations which are contaminated by losses in the system. Furthermore, the phase insensitive nature of our system makes it possible to extend the configuration to a large number of beams, as it avoids the phase locking required by linear beam-splitting method.

J. J. would like to thank Claude Fabre, Luis Orozco, Olivier Pfister, Paul Lett, Ping Koy Lam, Raphael Pooser, Nicolas Treps, and Travis Brannan for useful discussions. This work was supported by the National Basic Research Program of China (973 Program) under Grant No. 2011CB921604, the National Natural Science Foundation of China under Grants No. 11374104, No. 10974057, and No. 11234003, the SRFDP (20130076110011), the Program for Professor of Special Appointment (Eastern Scholar) at Shanghai Institutions of Higher Learning, the Program for New Century Excellent Talents in University (NCET-10-0383), the Shu Guang project supported by Shanghai Municipal Education Commission and Shanghai Education Development Foundation (11SG26), the Shanghai Pujiang Program under Grant No. 09PJ1404400, the Scientific Research Foundation for the Returned Overseas Chinese Scholars, State Education Ministry.

\nocite{*}

  *To whom all correspondence should be addressed. jtjing@phy.ecnu.edu.cn

\providecommand{\noopsort}[1]{}\providecommand{\singleletter}[1]{#1}%

\end{document}